\documentclass[11pt]{article}
\usepackage{a4}
\usepackage{times}
\usepackage{graphicx}
\usepackage{color}
\begin{document}
\title{Improving power posterior estimation of statistical evidence}
\author{
Nial Friel\thanks{School of Mathematical Sciences,
University College Dublin, Belfield, Dublin 4, Republic of Ireland;
Email: nial.friel@ucd.ie},
Merrilee Hurn\thanks{Department of Mathematical Sciences,
University of Bath, Bath, BA2 7AY, UK;
Email: M.A.Hurn@bath.ac.uk} and
Jason Wyse\thanks{School of Computer Science and Statistics,
Trinity College Dublin, College Green, Dublin 2, Republic of Ireland; Email: wyseja@scss.tcd.ie}
}

\maketitle

\begin{abstract}
\noindent
 The statistical evidence (or marginal likelihood) is a key quantity in Bayesian statistics, allowing one to assess
 the probability of the data given the model under investigation. This paper focuses on refining the power posterior 
 approach to improve estimation of the evidence. The power posterior method involves transitioning from the prior to 
 the posterior by powering the likelihood by an inverse temperature. In common with other tempering algorithms, the power 
 posterior involves some degree of tuning. The main contributions of this article 
 are twofold -- we present a result from the numerical analysis literature which can reduce the bias in the estimate 
 of the evidence by addressing the error arising from numerically integrating across the inverse temperatures. We also tackle 
 the selection of the inverse temperature ladder, applying this approach additionally to the Stepping Stone sampler estimation of evidence.
A key practical point is that both of these innovations incur virtually no extra cost. 

\end{abstract}

\section{Introduction}

The statistical evidence (sometimes called the marginal likelihood or integrated likelihood) is a vital quantity in 
Bayesian statistics for the comparison of models, $m_1,\dots,m_l$. Under the Bayesian paradigm we
consider the posterior distribution
\begin{equation}
 p(\theta_i,m_i|y) \propto p(y|\theta_i,m_i)p(\theta_i|m_i)p(m_i),\;\; \mbox{for}\;\; i=1,\dots,l,
\end{equation}
for data $y$ and parameters $\theta_i$ within model $m_i$, where $p(\theta_i|m_i)$ denotes the prior distribution for 
parameters within model $m_i$ and where $p(m_i)$ denotes the prior model probability. 
The evidence for data $y$ given model $m_i$ arises as the normalising constant of the posterior distribution within model $m_i$,
\begin{equation}
 p(\theta_i|y,m_i) \propto p(y|\theta_i,m_i)p(\theta_i|m_i),
\end{equation}
and thus results from integrating the un-normalised posterior across the $\theta_i$ parameter space,
\begin{equation}
 p(y|m_i) = \int_{\theta_i} p(y|\theta_i,m_i)p(\theta_i|m_i)\; d\theta_i.
\end{equation}
This of course assumes that the prior distribution for $\theta_i$ is proper. 
The marginal likelihood is often then used to calculate Bayes factors when one wants to compare two competing 
models, $m_i$ and $m_j$,
\begin{equation}
 BF_{ij} = \frac{p(y|m_i)}{p(y|m_j)} = \frac{p(m_i|y)}{p(m_j|y)} \frac{p(m_j)}{p(m_i)}.
\end{equation}
Here, $p(m_i|y)$ is the posterior probability for model $m_i$ and it can be evaluated, using the evidence
for each of the collection of models under consideration,
\begin{equation}
 p(m_i|y) \propto p(y|m_i) p(m_i),\;\; \mbox{for}\;\; i=1,\dots,l.
\end{equation}

Estimation of the evidence is a non-trivial
task for most statistical models and there has been considerable effort in the literature to find algorithms and methods 
for this purpose. Laplace's method (Tierney and Kadane, 1986)\nocite{tierney1986accurate} is an early approach and very widely used. 
Other notable and popular approaches include Chib's method (Chib 1995)\nocite{chib1995marginal}, annealed importance sampling 
(Neal 2001)\nocite{NealAnnealedImportance}, nested sampling (Skilling 2006)\nocite{skilling2006nested}, bridge sampling 
(Meng and Wong, 1996)\nocite{meng1996simulating} and power posteriors (Friel and Pettitt, 2008)\nocite{friel2008marginal} which 
is the focus of this paper. For a recent review and perspective on these and other methods, see Friel and 
Wyse (2012)\nocite{friel2011estimating}.

This paper is organised as follows. Section~\ref{sec:pp} outlines the power posterior method, and the approach we
propose to improve estimation of the evidence. Section~\ref{sec:examples} illustrates the potential gain from 
implementing the methodology which we propose. We offer some conclusions in Section~\ref{sec:conclusions}.

\section{The power posterior approach}
\label{sec:pp}

In what follows we will drop the explicit conditioning on model $m_i$
for notational simplicity.
We follow the notation of Friel and Pettitt (2008)\nocite{friel2008marginal} and denote the power posterior by
\begin{eqnarray}
\label{eqn:pp}
p_t(\theta| y ) & \propto & p(y|\theta)^t p(\theta) ,\ t \in [0,1] \\
\label{eqn:pp2}
\mbox{with } z(y|t) & = & \int_\theta p(y|\theta)^t p(\theta) d\theta.
\end{eqnarray}
where $t \in [0,1]$ is thought of as an inverse temperature, which has the effect of tempering the likelihood, whereby at 
the extreme ends of the inverse temperature range, $p_0(\theta|y)$ and $p_1(\theta|y)$ correspond to the prior and 
posterior, respectively. The power posterior estimator for the evidence relies on noting that
\begin{eqnarray} 
\frac{d}{dt} \log( z(y|t) ) & = & \frac{1}{z(y|t)} \frac{d}{dt} z(y|t) \nonumber\\
& = & \frac{1}{z(y|t)} \int_\theta \frac{d}{dt} p(y|\theta)^t p(\theta) d\theta \nonumber \\
& = & \frac{1}{z(y|t)} \int_\theta p(y|\theta)^t \log(p(y|\theta)) p(\theta) d\theta \nonumber \\
& = & \int_\theta \frac{p(y|\theta)^t p(\theta)}{z(y|t)} \log(p(y|\theta)) d\theta \nonumber
\\
& = & \mathbf{E}_{\theta|y,t} \log(p(y|\theta)).
\end{eqnarray} 
As a result
\begin{eqnarray}
\int_0^1 \mathbf{E}_{\theta|y,t} \log(p(y|\theta)) dt & = & 
\left [\ \log(z(y|t) )\ \right ]_0^1 \nonumber \\
& = & \log(z(y|t=1)) \mbox{ (assuming that the prior is normalised)}
\label{eqn:pp_evidence}
\end{eqnarray} 
which is the log of the desired marginal likelihood.

In practice the inverse temperature range is discretised as $0=t_0<t_1,\dots,t_n=1$ to form an estimator based on (\ref{eqn:pp_evidence}).
For each $t_i$, a sample from $p(\theta|y,t_i)$ can be used to estimate $\mathbf{E}_{\theta|y,t_i} \log(p(y|\theta))$. Finally, a 
trapezoidal rule is used to approximate 
\begin{equation}
\log p(y) \approx \sum_{i=1}^n (t_{i}-t_{i-1}) \left( 
\frac{ \mathbf{E}_{\theta|y,t_{i-1}} \log(p(y|\theta)) + \mathbf{E}_{\theta|y,t_i} \log(p(y|\theta)) }{2} \right). 
\label{eqn:quadrature}
\end{equation}
Discretising $t$ introduces an approximation into this method and 
the two goals of this paper are to reduce the bias in the power posterior estimation method due to the approximation 
and also to find an adaptive method for choosing the inverse temperatures (which we also refer to as rungs, following a common analogy of an inverse temperature ladder between prior and posterior).
For both of these we will exploit the fact that the gradient of the expected log deviance curve equals its variance, as we
now outline.

Differentiating $\mathbf{E}_{\theta|y,t} \log(p(y|\theta))$ with
respect to $t$ yields
\begin{eqnarray}
\frac{d}{dt} \mathbf{E}_{\theta|y,t} \log(p(y|\theta)) & = & 
\int_\theta \log(p(y|\theta)) \frac{d}{dt} p_t (\theta | y ) d\theta \nonumber \\
& = & \int_\theta \log(p(y|\theta)) \left [ \log(p(y|\theta)) 
- \frac{1}{z(y|t)} \frac{d}{dt} z(y|t) \right ] p_t(\theta | y ) d\theta \nonumber \\
& = & \int_\theta \log(p(y|\theta)) \left [ \log(p(y|\theta)) 
- \frac{d}{dt} \log(z(y|t)) \right ] p_t(\theta | y ) d\theta \nonumber \\
& = & \mathbf{E}_{\theta|y,t} \log(p(y|\theta))^2 - 
(\mathbf{E}_{\theta|y,t} \log(p(y|\theta)))^2 \nonumber \\
& = & \mathbf{V}_{\theta|y,t}(\log(p(y|\theta)))
\label{eqn:gradvar}
\end{eqnarray}
where $\mathbf{V}_{\theta|y,t}(\log(p(y|\theta)))$ denotes the variance of the log deviance at inverse temperature $t$.

\subsection{Reducing the bias by improving the numerical integration}
Equation~(\ref{eqn:gradvar}) immediately provides two useful pieces of information.
First, the curve which we wish to integrate numerically is (strictly) increasing.
Secondly, we can improve upon the standard trapezium rule used to numerically integrate the expected log deviance by incorporating 
derivative information at virtually no extra computational cost (the cost merely of calculating the variance of a set of simulations for fixed $t$).
We do this by using the corrected trapezium rule which comes from an error analysis of the standard trapezium rule, see for example Atkinson and Han (2004)\nocite{atkinson2004han}, Section 5.2;
when integrating a function $f$ between points $a$ and $b$
\begin{equation}
\int_a^b f(x) dx = (b-a)\left[ \frac{f(b)+f(a)}{2}\right] - \frac{(b-a)^3}{12}
f^{\prime\prime}(c)
\end{equation}
where $c$ is some point in $[a,b]$.
The first term of the right hand side of this equation is the usual trapezium rule and the second can be approximated using 
\begin{eqnarray}
f^{\prime\prime}(c) & \approx & \frac{ f^\prime(b) - f^\prime(a)}{b-a} \nonumber \\
\mbox{so that }\int_a^b f(x) dx & \approx & (b-a)\left[ \frac{f(b)+f(a)}{2}\right] - \frac{(b-a)^2}{12} \left [f^\prime(b) - f^\prime(a) \right ].
\label{eqn:approx}
\end{eqnarray}
This latter form motivates the corrected trapezium rule which for unequally spaced x-axis points, taken together with the information derived above regarding the derivative of the log deviance gives
\begin{eqnarray}
 \log(z(y|t=1)) & \approx & \sum_{i=0}^{n-1} (t_{i+1}-t_i) \left [ \frac{\mathbf{E}_{\theta|y,t_i}\log(p(y|\theta)) + \mathbf{E}_{\theta|y,t_{i+1}}\log(p(y|\theta))}{2} \right ] \nonumber \\
& & - \sum_{i=0}^{n-1} \frac{(t_{i+1}-t_i)^2}{12} \left [ \mathbf{V}_{\theta|y,t_{i+1}}(\log(p(y|\theta))) - \mathbf{V}_{\theta|y,t_{i}}(\log(p(y|\theta)) ) \right ]
\label{eqn:modify}
\end{eqnarray}
where both the expectations $\{\mathbf{E}_{\theta|y,t_{i}}\log(p(y|\theta))\}$ and variances $\{\mathbf{V}_{\theta|y,t_{i}}(\log(p(y|\theta)))\}$ are to be estimated using MCMC runs at a number of values of $t_i$.
We will refer to the algorithm incorporating this correction as the modified power posterior.

\subsection{Adaptive choice of the inverse temperature placement}
\label{sec:adap}

The next question which arises is how to choose the $\{t_i\}$ between $t_0=0$ and $t_n=1$.
Friel and Pettitt (2008)\nocite{friel2008marginal} find that setting $t_i = (i/n)^5$ performs well. We refer to this as the powered fraction (PF) schedule.
Lartillot and Philippe~(2006)\nocite{lartillot2006computing} discuss very similar ideas in the phylogenetics literature, although using Simpson's rule for the numerical integration; they use equally spaced inverse temperatures between 0 and 1.

Here we will only consider the discretisation error associated with the numerical integration, rather than the stochastic error arising with sampling from the different $p_{t_i}(\theta|y)$.
Calderhead and Girolami (2009) \nocite{calderhead2009estimating} show that this discretisation error depends upon the Kullback-Leibler distance between successive $p_{t_i}(\theta|y)$.
Lefebvre, Steele and Vandal (2010)\nocite{Lef08} also consider a symmetrised Kullback-Leibler divergence in picking optimal schedules for path sampling.
At first glance the Kullback-Leibler distance does not seem a particularly tractable quantity to manipulate.
However, these papers and Behrens, Friel and Hurn (2012)\nocite{behrens2012tuning} all note that, in the notation of this paper,
\begin{equation}
\sum_{i=0}^{n-1} ( KL[p_{t_i}(\theta|y) ,p_{t_{i+1}}(\theta|y)] +
KL[p_{t_{i+1}}(\theta|y) ,p_{t_{i}}(\theta|y)] )
= 2 S_n(t_0,\ldots,t_n)
\end{equation}
where $KL$ denotes the Kullback-Leibler distance and
\begin{equation}
S_n(t_0,\ldots,t_n) = \sum_{i=0}^{n-1} (t_{i+1}-t_{i})
\mathbf{E}_{\theta|y,t_{i+1}}\log(p(y|\theta)) -
\sum_{i=0}^{n-1} (t_{i+1}-t_{i})
\mathbf{E}_{\theta|y,t_{i}}\log(p(y|\theta)) .
\end{equation}
$S_n$ can be interpreted graphically as the sum of the rectangular areas between a lower and an upper approximation to the integral of $\mathbf{E}_{\theta|y,t_{i}}\log(p(y|\theta))$ between $t_0=0$ and $t_1=1$.
Behrens, Friel and Hurn (2012)\nocite{behrens2012tuning} use minimising $S_n$ as a rationale for choosing the inverse temperatures in tempered transitions.
We propose to use the same target in selecting the $\{t_i\}$ for power posteriors.
However, unlike in tempered transitions where the tuning forms a small part of the overall computational load, here the cost is almost exclusively the estimation of $\mathbf{E}_{\theta|y,t_{i}}\log(p(y|\theta))$ and its gradient.

We propose the following approach.
Begin by estimating the expected log deviance and its gradient at $t=1$ then $t=0$ (both points which are needed in all possible schemes).
Where to site the next $t$?
There is no analytic solution without knowing the curve and we do not want to use computational resources in performing a search for the optimal location.
Instead we find the intersection of the two straight lines defined by our current knowledge of the curve:
If the estimated function and gradient at $t_k$ are denoted by $\hat{f}_k$ and $\hat{V}_k$ respectively, and those at $t_{k+1}$ by $\hat{f}_{k+1}$ and $\hat{V}_{k+1}$, we set the new point to be
\begin{equation}
t = \frac{\hat{f}_{k+1}-\hat{f}_k+t_k\hat{V}_k-t_{k+1}\hat{V}_{k+1}}{\hat{V}_k-\hat{V}_{k+1}} .
\end{equation}
If this intersecting point is outside the interval $[t_k,t_{k+1}]$, it suggests there is some sort of inflection within the interval and instead we use a simple weighted average.
\begin{equation}
t = t_k + \frac{\hat{V}_{k+1}}{\hat{V}_k+\hat{V}_{k+1}} (t_{k+1} - t_k).
\end{equation}
This now gives us two rectangular contributions to $S_n$.
We identify the larger of these terms and locate the next point in the corresponding interval, and so on iteratively.
This scheme will almost certainly not identify the optimal placing of the $n-1$ interior rungs.
However it is quick, cheap and intuitively reasonable.
(In practice, Monte Carlo error can mean that the function is not increasing and so the criterion is changed to picking the interval with the largest absolute contribution to $S_n$.
In the event of picking such an interval, we set the new $t$ to be the midpoint of the interval reflecting the significant levels of uncertainty.)

\subsection{Computational details}
\label{sec:details}
We use sequential runs of MCMC, beginning with sampling from the posterior, $t=1$. 
From each MCMC run we store the estimated expected log deviance, its estimated gradient and the values at the last iteration of the run.
These final values are used as starting points for runs at a smaller value of $t$.
In the case of the power fraction placements where the inverse temperatures are deterministic, this means that the run at $t_k$ is initialised with the values from $t_{k+1}$.
In the adaptive placements, the run at each new $t$ is initialised with the values of the currently closest larger $t_k$.

\section{Examples}
\label{sec:examples}

We present three examples,
the first two of which were included in the review paper by
Friel and Wyse (2012)\nocite{friel2011estimating} where the performance of power posteriors was compared to some other existing methods.
The first is a non-nested linear regression comparison for which the
marginal likelihoods can be calculated analytically.
Here our experiments concentrate on the effects of the modified integration rule and the adaptive placement of the inverse temperatures.
Example 2 is a larger problem, choosing between two logistic regression models, for which an analytic solution is not possible;
we use this example to compare the improved algorithm with the Stepping Stone sampling approach of Xie {\it et al} (2011)\nocite{xie2011improving}.
The final example is by far the largest and exhibits the most interestingly shaped $\mathbf{E}_{\theta|y,t}\log(p(y|\theta))$, the focus here is on the stability of the approach to poor estimation of the gradient.

\subsection{Example 1: Radiata pine}
\label{sec:radiata}
The first example compares two linear regression models for the 
Radiata pine data originally in Williams (1959)\nocite{will59}.
The response variable here is the maximum compression strength parallel to the grain, $y_i$, while the predictors are density, $x_i$, or density adjusted for
resin content, $z_i$, for $n = 42$ specimens of radiata pine.
Two possible Gaussian linear regression models are considered;
\[
\begin{array}{lccc}
\mbox{Model 1: } & y_i = \alpha + \beta(x_i - \bar{x}) + \epsilon_i, & \epsilon_i \sim N(0,\tau^{-1}), &i = 1,\dots,n,\\
\mbox{Model 2:} & y_i = \gamma + \delta(z_i - \bar{z}) + \eta_i, & \eta_i \sim N(0,\lambda^{-1}), & i = 1,\dots,n.\\
\end{array}
\]
Priors are chosen to match the analyses of Friel and Wyse (2012)\nocite{friel2011estimating} (baring a notational factor of 2).
The regression parameters $(\alpha,\beta)^T$ and $(\gamma,\delta)^T$ 
were taken to be Normally distributed with
mean $(3000,185)^T$ and precision $\tau Q_0$ and $\lambda Q_0$ respectively
where $Q_0 = \mbox{diag}(r_0,s_0)$. The values of $r_0$ and $s_0$ were fixed to be 0.06 and 6. A gamma prior with shape
$a_0 = 3$ and rate $b_0 = 2\times300^2$ was assumed for both $\tau$ and $\lambda$.

We consider estimating the evidence using $n=10$, 20, 50 or 100 rungs in the tempering scheme where the rungs are chosen either according to the powered fraction (PF) rule or adaptively following the heuristic in Section~\ref{sec:adap}.
The parameters at all levels are updated using the Gibbs sampler, with 10000 iterations at each rung, discarding the first fifth of these as burn in.
To quantify the performances, the bias, standard deviation and Root Mean Square Error (RMSE) are estimated by performing 100 replicates of the four schemes at 10, 20, 50 and 100 rungs.
The results are given in Table~\ref{tab:pine}.
Comparing the first column with the second, and the third with the fourth, we can see that the modified power posterior approach dominates the usual one in all cases.
For small numbers of rungs, where the RMSE is dominated by the bias term, this effect is particularly dramatic.
The standard error is not greatly affected by the choice of scheme, although in this example the smallest values all occur for modified schemes.
As the number of rungs increases, the RMSE becomes more affected by the standard error than the bias.
\begin{table}[]
\begin{center}
\begin{tabular}{| l|l|l|r|r|r|r|}
\hline
& & Power posterior: & Standard & Modified & Standard & Modified \\
&  Rungs: & & PF & PF & Adaptive & Adaptive \\
\hline \hline
MODEL 1 & $n=10$ & Bias & -0.6569 & 0.0970 & -0.4363 & {\bf0.0434}  \\
&  & Standard error & 0.0246 & {\bf0.0196} & 0.0216 & 0.0199  \\
&  & RMSE & 0.6573 & 0.0990 & 0.4369 & {\bf0.0478}  \\
\hline
& $n=20$ & Bias & -0.1628 & {\bf0.0044} & -0.1128 & 0.0057  \\
&  & Standard error & 0.0161 & {\bf0.0153} & 0.0163 & 0.0154  \\
&  & RMSE & 0.1636 & {\bf0.0160} & 0.1140 & 0.0164  \\
\hline
& $n=50$ & Bias & -0.0258 & {\bf0.0000} & -0.0251 & -0.0041  \\
&  & Standard error & 0.0098 & {\bf0.0097} & 0.0104 & 0.0101  \\
&  & RMSE & 0.0276 & {\bf0.0097} & 0.0272 & 0.0109  \\
\hline
& $n=100$ & Bias & -0.0059 & {\bf0.0005} & -0.0101 & -0.0041  \\
&  & Standard error & 0.0086 & 0.0085 & 0.0080 & {\bf0.0079}  \\
&  & RMSE & 0.0104 & {\bf0.0086} & 0.0129 & 0.0089  \\
\hline
\hline
MODEL 2 & $n=10$ & Bias & -0.6354 & 0.1012 & -0.4262 & {\bf0.0336}  \\
&  & Standard error & 0.0247 & {\bf0.0197} & 0.0253 & 0.0228  \\
&  & RMSE & 0.6359 & 0.1031 & 0.4269 & {\bf0.0406}  \\
\hline
& $n=20$ & Bias & -0.1585 & 0.0042 & -0.1116 & {\bf0.0029}  \\
&  & Standard error & 0.0170 & 0.0160 & 0.0152 & {\bf0.0141}  \\
&  & RMSE & 0.1594 & 0.0165 & 0.1126 & {\bf0.0144}  \\
\hline
& $n=50$ & Bias & -0.0249 & {\bf0.0002} & -0.0241 & -0.0038  \\
&  & Standard error & 0.0106 & 0.0106 & 0.0103 & {\bf0.0101}  \\
&  & RMSE & 0.0270 & {\bf0.0106} & 0.0262 & 0.0108  \\
\hline
& $n=100$ & Bias & -0.0073 & {\bf-0.0011} & -0.0085 & -0.0027  \\
&  & Standard error & 0.0084 & 0.0084 & 0.0062 & {\bf0.0061}  \\
&  & RMSE & 0.0112 & 0.0085 & 0.0105 & {\bf0.0066}  \\
\hline
\end{tabular}
\caption{Estimated bias, standard error and Root Mean Square Error for the two Radiata pine models using different numbers of rungs and different schemes (the standard power posterior vs the modified version, the Power Fraction scheme vs the adaptive scheme).
For each model and number of rungs, the minimum absolute bias, standard error and RMSE are highlighted in bold.
There are 100 replicates used for estimation, all use 10000 iterations at each rung discarding the first fifth as burn in.}
\label{tab:pine}
\end{center}
\end{table}

\begin{figure}[tbp] 
\begin{center}
\includegraphics[width=15cm, height=21cm]{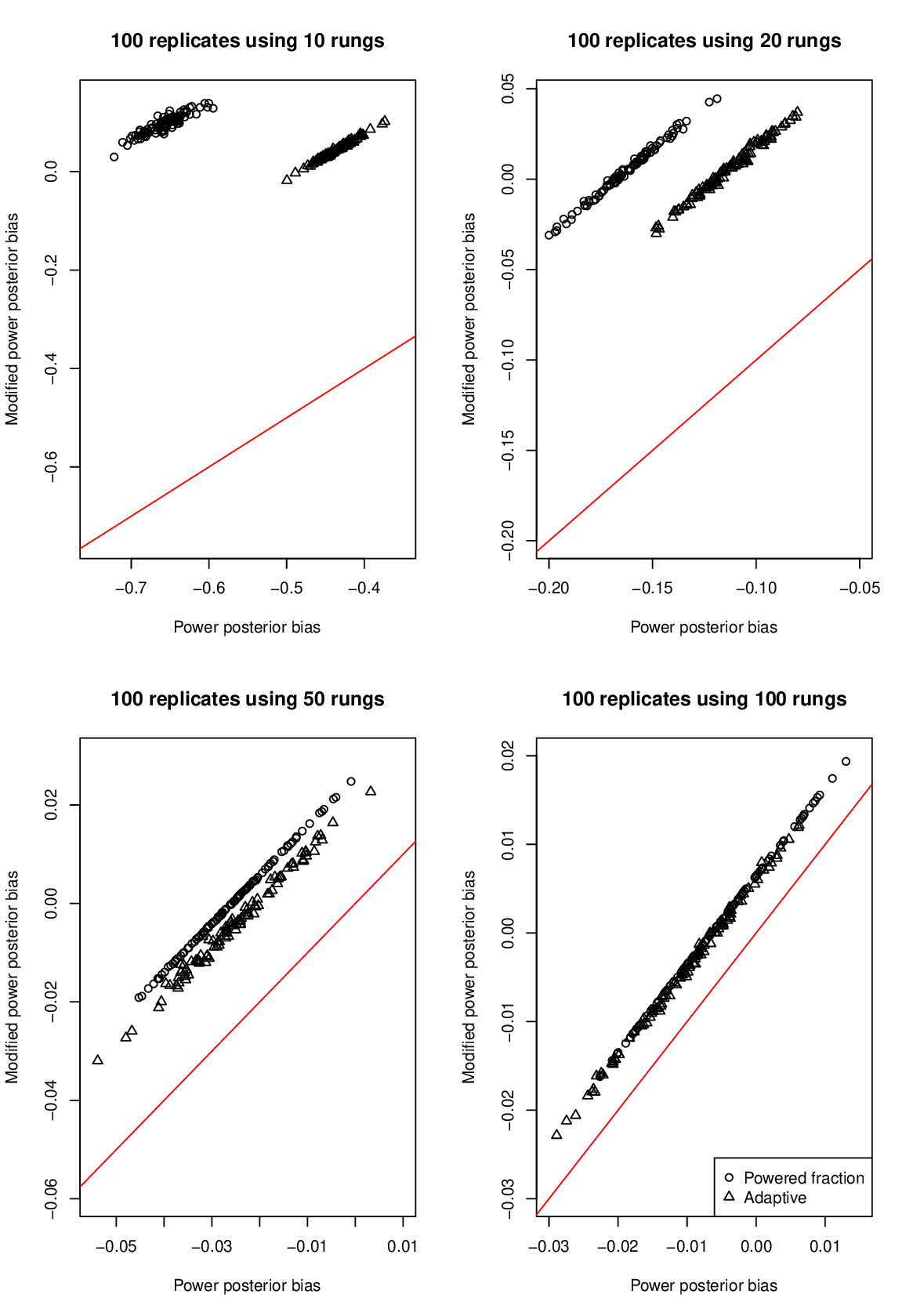}
\caption{The 100 observed biases for Model 1 for the Radiata data using either PF or adaptive spacings plotted as standard vs modified pairs. The vertical distances from the red $y=x$ lines indicate the size of the correction in the modified power posterior.}
\label{fig:pine1}
\end{center}
\end{figure}

Comparing the first column of Table~\ref{tab:pine} with the third, we can isolate the effect of the adaptive placement.
This is also most noticeable at small numbers of rungs, perhaps unsurprisingly given that the placing of an individual $t_i$ makes less difference as their total number increases.
To help to visualise the effect of both the correction and the adaptive placing of inverse temperatures, Figure~\ref{fig:pine1} plots the 100 observed biases pairwise for the standard vs modified power posterior.
Points are plotted for both the PF spacing and the adaptive spacings, and for each number of rungs.
How much change occurs for each standard power posterior under the additive modification given by Equation~(\ref{eqn:modify}) can be seen as the vertical distance between the point and the red $y=x$ line.
A few points are immediately clear.
Firstly the correction is fairly well estimated (the sets of points are roughly parallel to the $y=x$ line, even in the $n=10$ case; this is a function of the MCMC run lengths).
Second, there is an interesting difference between the adaptive and the PF versions, less correction is needed for the adaptive schedule, that is, it is already doing a better job of linearly approximating the function for the numerical integration.
Thirdly, as was already observed in Table~\ref{tab:pine}, as the number of rungs increases, the differences between the two types of spacings become less pronounced.

Given the good reductions in bias seen in Table~\ref{tab:pine}, it is important to ask how much extra time is required.
To assess this, a total of 20 runs for Model 1 using 10000 iterations and 100 inverse temperatures were timed.
Four versions of the algorithm were considered, corresponding to Table~\ref{tab:pine}. 
All the coding was in R and times are given relative to the PF standard power posterior version:
\[
\begin{array}{cccc}
\mbox{PF standard} & \mbox{PF modified} & \mbox{Adaptive standard} & \mbox{Adaptive modified} \\
1.0000 & 1.0083 & 1.0076 & 1.0121 \\
\end{array}
\]
What we see is that the adaptive selection of inverse temperatures and the correction term in the numerical integration come at an acceptably small computational cost.
Given this, our recommendation would be to use the modified integration rule when employing power posteriors.
The benefits of the adaptive placement are less dramatic except at small $n$ but as it provides a completely automatic way to choose the inverse temperatures for very little cost, we would also recommend it.

\subsection{Example 2: Pima indians}
\label{sec:pima}
We turn next to the Pima Indian example considered by Friel and Wyse (2012)\nocite{friel2011estimating}, originally described by Smith {\it et al}~(1988)\nocite{smith1988using} .
These data record diabetes incidence and possible disease indicators for
$n = 532$ Pima Indian women aged over 20. 
The seven possible disease indicators are
the number of pregnancies (\texttt{NP}), plasma glucose concentration (\texttt{PGC}), diastolic blood pressure (\texttt{BP}), triceps
skin fold thickness (\texttt{TST}), body mass index (\texttt{BMI}), diabetes pedigree function (\texttt{DP}) and age (\texttt{AGE}),
with all these covariates standardised.

The model assumed for the observed diabetes incidence, $y=(y_1,\ldots,y_n)$, is 
\begin{equation}
p(y|\theta) = \prod_{i=1}^n p_i^{y_i}(1-p_i)^{1-y_i}
\end{equation}
where $p_i$ is the probability of incidence for person $i$, and $p_i$ is related to the $i^{th}$ person's covariates and a constant term, denoted by
$x_i = (1,x_{i1},\dots, x_{id})^{T}$, and the parameters, $\theta = (\theta_0,\theta_1,\dots,\theta_d)^{T}$, by
\begin{equation}
\log\left(\frac{p_i}{1-p_i}\right) = \theta^{T} x_{i}
\end{equation}
where $d$ is the number of explanatory variables. An independent multivariate Gaussian prior is assumed for 
$\theta$, with mean zero and 
non-informative precision of $\tau=0.01$, so that
\begin{equation}
p(\theta) = (2\pi)^{-(d+1)/2} \tau^{(d+1)/2} \exp\left\{-\frac{\tau}{2}\theta^{T}\theta \right\}.
\end{equation}

\begin{figure}[tbp] 
\begin{center}
\includegraphics[width=15cm, height=21cm]{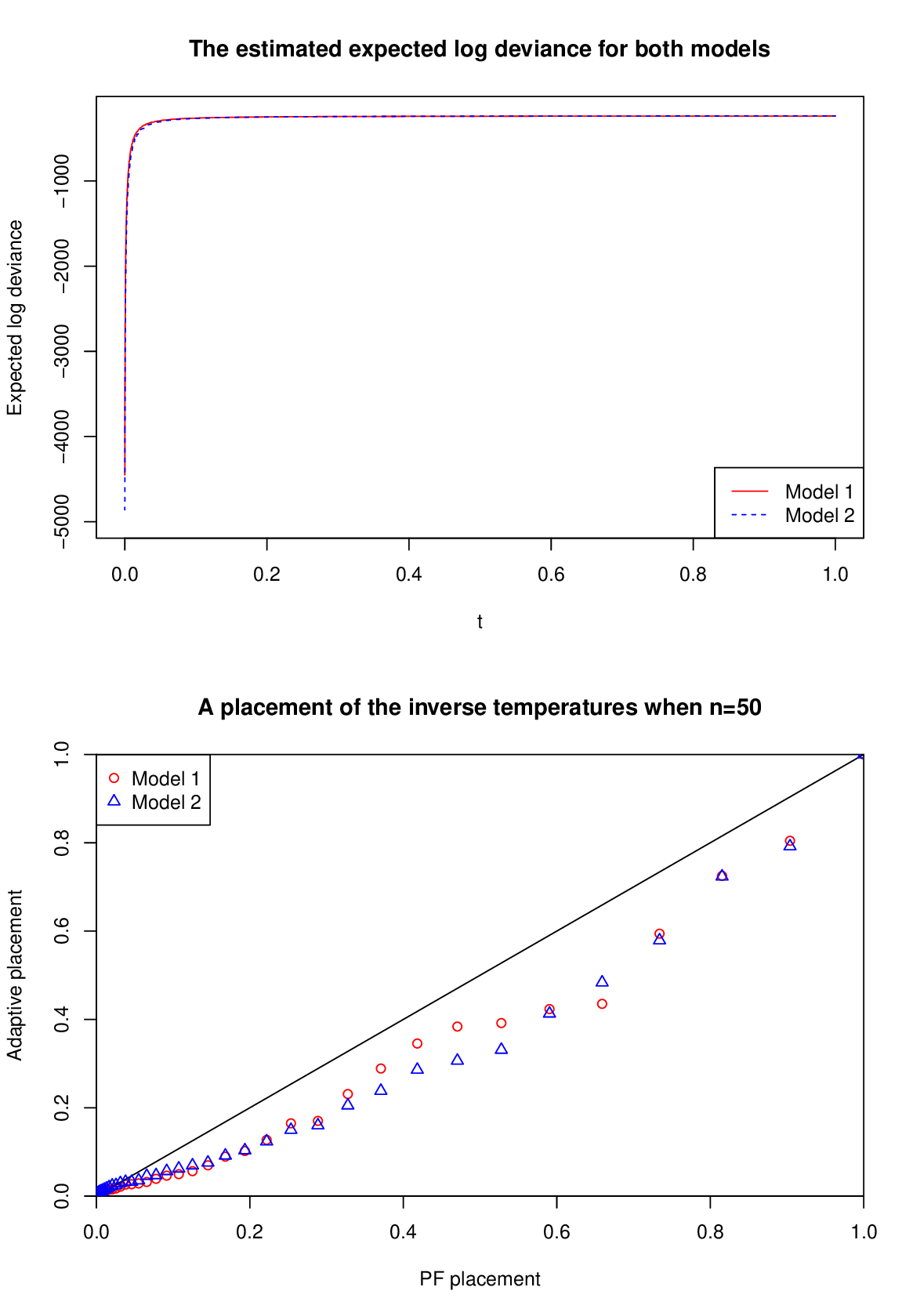}
\caption{Top: The expected log deviance curves for the Pima models (estimated using 200 PF inverse temperatures with the modified power posterior each with 20000 iterations, discarding the first fifth as burn in).
Bottom: A realisation of the adaptive temperature placements for the two models plotted against the PF scheme when $n=50$.}
\label{fig:pima1}
\end{center}
\end{figure}

A long reversible jump run (Green, 1995\nocite{green1995reversible}) revealed the two models with the highest posterior probability:
\begin{center}
$\begin{array}{lll}
\mbox{Model 1:} & \texttt{logit(p)} $=$ & \texttt{1} $+$ \texttt{NP} $+$ \texttt{PGC} $+$ \texttt{BMI} $+$ \texttt{DP} \\
\mbox{Model 2:} & \texttt{logit(p)} $=$ & \texttt{1} $+$ \texttt{NP} $+$ \texttt{PGC} $+$ \texttt{BMI} $+$ \texttt{DP} $+$ \texttt{AGE} \\
\end{array}$
\end{center}
For these two models, the power posterior is not amenable to the Gibbs sampler and so a Metropolis update is used instead. 
This raises the problem of proposal scaling at the different inverse temperatures.
Since both the correction and the adaptive inverse temperature placements assume good estimates of the variance of the log deviance, mixing is an important issue; we return to the question of robustness to the quality of these estimates in Section~\ref{sec:galaxy}.
As an alternative to the sampler used in Friel and Wyse (2012)\nocite{friel2011estimating}, we work with a joint update of all the model parameters using a multivariate Normal proposal centred at the current value and with diagonal variance matrix, entries $\min( 0.01/t , 1/\tau )$ where $t$ is the inverse temperature and $\tau$ is the precision of the prior.
The top panel of Figure~\ref{fig:pima1} shows the estimated log deviance curves for these two competing models using this sampler in the modified power posterior with 200 PF inverse temperatures each with 20000 iterations, discarding the first fifth as burn in.
The bottom panel of the same figure shows two realisations of the adaptive placements, one for each model, against the PF scheme with $n=50$.

Unlike in the Radiata example where the evidence can be evaluated analytically, here some estimation is required.
Friel and Wyse (2012)\nocite{friel2011estimating} use the Laplace approximation of the log evidence (Tierney and Kadane, 1986\nocite{tierney1986accurate}) as the ``benchmark'' in assessing bias.
However this is not necessarily very accurate and so we replace the Laplace approximation by a very long run (2000 rungs, each using 20000 iterations) of the power posterior approach;
the estimates we get are $-257.2342$ and $-259.8519$ for Models 1 and 2 respectively as opposed to the Laplace approximations of $-257.2588$ and $-259.8906$.
We will use these values in a comparison of performance, with and without tuning, of power posteriors and 
another approach to estimating the evidence, the recent Stepping Stone sampler of Xie {\it et al} (2011)\nocite{xie2011improving}.

Based on importance sampling, the Stepping Stones sampler uses the same idea of powered posteriors, Equation~(\ref{eqn:pp}), as a series of intermediate distribution, the ``stepping stones'', between the prior and the posterior, but utilising them as importance distributions rather than performing numerical integration.
It generates samples from each of the power posteriors from $t_0=0$ up to $t_{n-1}$, estimating the ratio of consecutive normalising constants $r_{k}=z(y|t_{k+1})/z(y|t_{k})$, where $z(y|t)$ is given by Equation~(\ref{eqn:pp2}), using the $t_k$ power posterior as an importance distribution:
\begin{equation}
\hat{r}_k = \frac{1}{m} \sum_{i=1}^m p(y|\theta^{(i)})^{t_{k+1}-t_k},\ k=0,\ldots,n-1
\label{eqn:ss}
\end{equation}
where $m$ is the number of MCMC samples post burn in and the corresponding sampled values $\{\theta^{(i)}\}$ are drawn from $p_{t_k}(\theta| y)$. 
Assuming that the prior is normalised, the final estimate of the evidence, $z(y| t_n=1)$, is the product of these $n$ independent estimates, $\prod_{k=0}^{n-1} \hat{r}_k$. 
The Stepping Stone sampler is unbiased for the marginal likelihood, although slightly biased for the log marginal likelihood. 
Xie {\it et al} (2011)\nocite{xie2011improving} compare its performance in estimating the log marginal likelihood to that of power posteriors,
finding that for a well chosen inverse temperature placement or for a large number of inverse temperatures, power posteriors and the Stepping Stone approach are comparable, with the latter slightly outperforming the former, but when there are few inverse temperatures or badly placed ones, power posterior estimates perform relatively poorly.

\begin{table}[]
\begin{center}
\begin{tabular}{| l|l|l|r|r|r|r|}
\hline
& & & Standard PP & Modified PP & Stepping Stone & Stepping Stone \\
&  Rungs:&  & PF & Adaptive & PF & Adaptive \\
\hline \hline
MODEL 1 & $n=10$ & Bias & -3.67946 & 0.64809 & {\bf-0.02251} & -0.03174 \\
&  & Standard error & 0.35152 & 0.25121 & 0.25666 & {\bf0.18835}  \\
&  & RMSE & 3.69621 & 0.69507 & 0.25765 & {\bf0.19101}  \\
\hline
& $n=20$ & Bias & -0.85666 & 0.03552 & 0.01973 & {\bf0.00865} \\
&  & Standard error & 0.23478 & 0.16722 & 0.18491 & {\bf0.14805}  \\
&  & RMSE & 0.88825 & 0.17095 & 0.18596 & {\bf0.14830}  \\
\hline
& $n=50$ & Bias & -0.13628 & -0.00767 & {\bf-0.00150} & 0.03435  \\
&  & Standard error & 0.13008 & {\bf0.09737} & 0.11322 & 0.09836  \\
&  & RMSE & 0.18840 & {\bf0.09767} & 0.11323 & 0.10418  \\
\hline
& $n=100$ & Bias & {\bf-0.00383} & 0.01845 & 0.02736 & 0.09785  \\
&  & Standard error & 0.10495 & 0.08248 & 0.09777 & {\bf0.08172}  \\
&  & RMSE & 0.10502 & {\bf0.08452} & 0.10153 & 0.12749  \\
\hline
\hline
MODEL 2 & $n=10$ & Bias & -4.16969 & 0.77087 & {\bf-0.00362} & 0.01845 \\
&  & Standard error & 0.33864 & 0.32020 & 0.28518 & {\bf0.25373} \\
&  & RMSE & 4.18342 & 0.91444 & 0.28521 & {\bf0.25440} \\
\hline
& $n=20$ & Bias & -0.94958 & 0.02703 & 0.03612 & {\bf-0.01686} \\
&  & Standard error & 0.25089 & 0.20139 & 0.19515 & {\bf0.18510} \\
&  & RMSE & 0.98216 & 0.20212 & 0.19847 & {\bf0.18587} \\
\hline
& $n=50$ & Bias & -0.11702 & -0.03084 & 0.02995 & {\bf0.01999} \\
&  & Standard error & 0.15566 & 0.12323 & 0.13724 & {\bf0.11467} \\
&  & RMSE & 0.19474 & 0.12418 & 0.14047 & {\bf0.11640} \\
\hline
& $n=100$ & Bias & {\bf-0.00384} & 0.02610 & 0.03580 & 0.11592 \\
&  & Standard error & 0.11019 & {\bf0.09285} & 0.10378 & 0.09334 \\
&  & RMSE & 0.11025 & {\bf0.09353} & 0.10978 & 0.14883 \\
\hline
\end{tabular}
\caption{Estimated bias, standard error and Root Mean Square Error for the two Pima Indian models using different numbers of rungs and different schemes (the standard power posterior with PF, the modified power posterior with adaptive rungs, and the Stepping Stone sampler with both PF and adaptive schemes).
There are 100 replicates used for the estimation, all use 10000 iterations at each rung discarding the first fifth as burn in.}
\label{tab:pima}
\end{center}
\end{table}

\begin{table}[]
\begin{center}
\begin{tabular}{| l|l|l|r|r|r|r|}
\hline
& & & Standard PP & Modified PP & Stepping Stone & Stepping Stone \\
&  Rungs:&  & PF & Adaptive & PF & Adaptive \\
\hline \hline
MODEL 1 & $n=10$ & Bias & -3.64242 & 0.72598 & {\bf0.01293} & 0.02300 \\
&  & Standard error & 0.10541 & 0.09356 & 0.07597 & {\bf0.06201}  \\
&  & RMSE & 3.64395 & 0.73198 & 0.07706 & {\bf0.06614}  \\
\hline
& $n=20$ & Bias & -0.87596 & 0.04038 & 0.01651 & {\bf0.01554} \\
&  & Standard error & 0.06776 & 0.05361 & 0.05348 & {\bf0.04788}  \\
&  & RMSE & 0.87858 & 0.06712 & 0.05597 & {\bf0.05034}  \\
\hline
& $n=50$ & Bias & -0.11517 & {\bf0.01508} & 0.02568 & 0.02021  \\
&  & Standard error &  0.04333 & 0.03735 & 0.03705 & {\bf0.03500}  \\
&  & RMSE & 0.12305 & {\bf0.04028} & 0.04508 & 0.04041  \\
\hline
& $n=100$ & Bias & -0.01262 & {\bf0.01133} & 0.02210 & 0.02138  \\
&  & Standard error & 0.02961 & 0.02511 &  0.02752 & {\bf0.02454}  \\
&  & RMSE & 0.03219 & {\bf0.02755} & 0.03529 & 0.03254  \\
\hline
\hline
MODEL 2 & $n=10$ & Bias & -4.17639 & 0.80148 & 0.03336 & {\bf0.02949} \\
&  & Standard error & 0.13585 & 0.10192 & 0.09553 & {\bf0.08534} \\
&  & RMSE & 4.17860 & 0.80794 & 0.10119 & {\bf0.09029} \\
\hline
& $n=20$ & Bias & -0.97693 & 0.05061 & 0.03903 & {\bf0.02079} \\
&  & Standard error & 0.08038 & 0.05887 & 0.06134 & {\bf0.05227} \\
&  & RMSE & 0.98024 & 0.07763 & 0.07270 & {\bf0.05626} \\
\hline
& $n=50$ & Bias & -0.13474 & {\bf0.01974} & 0.02745 & 0.02509\\
&  & Standard error & 0.04836 & 0.04230 & 0.04502 & {\bf0.03978} \\
&  & RMSE & 0.14316 & {\bf0.04668} & 0.05273 & 0.04703 \\
\hline
& $n=100$ & Bias & {\bf-0.00197} & 0.02507 & 0.03797 & 0.03753 \\
&  & Standard error & 0.03727 & 0.02933 & 0.03546 & {\bf0.02809} \\
&  & RMSE & {\bf0.03732} & 0.03859 & 0.05195 & 0.04688\\
\hline
\end{tabular}
\caption{Estimated bias, standard error and Root Mean Square Error for the two Pima Indian models using different numbers of rungs and different schemes (the standard power posterior with PF, the modified power posterior with adaptive rungs, and the Stepping Stone sampler with both PF and adaptive schemes).
There are 100 replicates used for the estimation, all use 100000 iterations at each rung discarding the first fifth as burn in (that is, ten times the run lengths and burn ins of Table~\ref{tab:pima}).}
\label{tab:longer}
\end{center}
\end{table}

Table~\ref{tab:pima} shows the estimated biases, standard errors and RMSEs for both models and for four possible schemes: standard power posteriors with PF spacings, modified power posteriors with adaptive spacings, Stepping Stone with PF spacings, Stepping Stone with adaptive spacings.
There are 10000 iterations at each rung discarding the first fifth as burn in.
Notice that for the Stepping Stone sampler, the samples from the posterior (i.e. $t_n=1$) are not required, other than indirectly to initialise the $t_{n-1}$ sample; other than that, the two approaches use exactly the same MCMC samples.
What we can see from Table~\ref{tab:pima} is that the Stepping Stone estimates have generally small biases independent of the number of rungs or their placement.
However something unexpected occurs for the Stepping Stones estimates for larger $n$ with adaptive spacings, in that their bias begins to increase.
This does not occur if the adaptively placed rungs from a power posterior run are used as deterministic rungs in a separate Stepping Stones run.
To explore this further, we ran additional experiments, replacing 10000 iterations at each rung (discarding the first 2000 as burn-in) by 100000 iterations (discarding the first 20000 as burn-in).
The results are presented in Table~\ref{tab:longer}.
What we see is that there is no longer a discrepancy between the bias of the Stepping Stones algorithm using the deterministic and the bias using the adaptive rung placements.
The fact that it is only the adaptive Stepping Stones biases which change noticeably with the increase in run length, suggests to us
that Stepping Stones is perhaps more sensitive than power posteriors to the fact that for adaptive placement of the rungs, the initialisation of the MCMC is not necessarily at the next largest of the final set of $\{t_i\}$ as it is with a deterministic placement.
Whilst it might seem odd to present unconverged results, the point we would like to make is that there was no hint of lack of convergence visually inspecting the MCMC chains and it is only the adaptive Stepping Stone for which this quiet lack of convergence has had an effect here.

\begin{figure}[tbp] 
\begin{center}
\includegraphics[width=15cm, height=20cm]{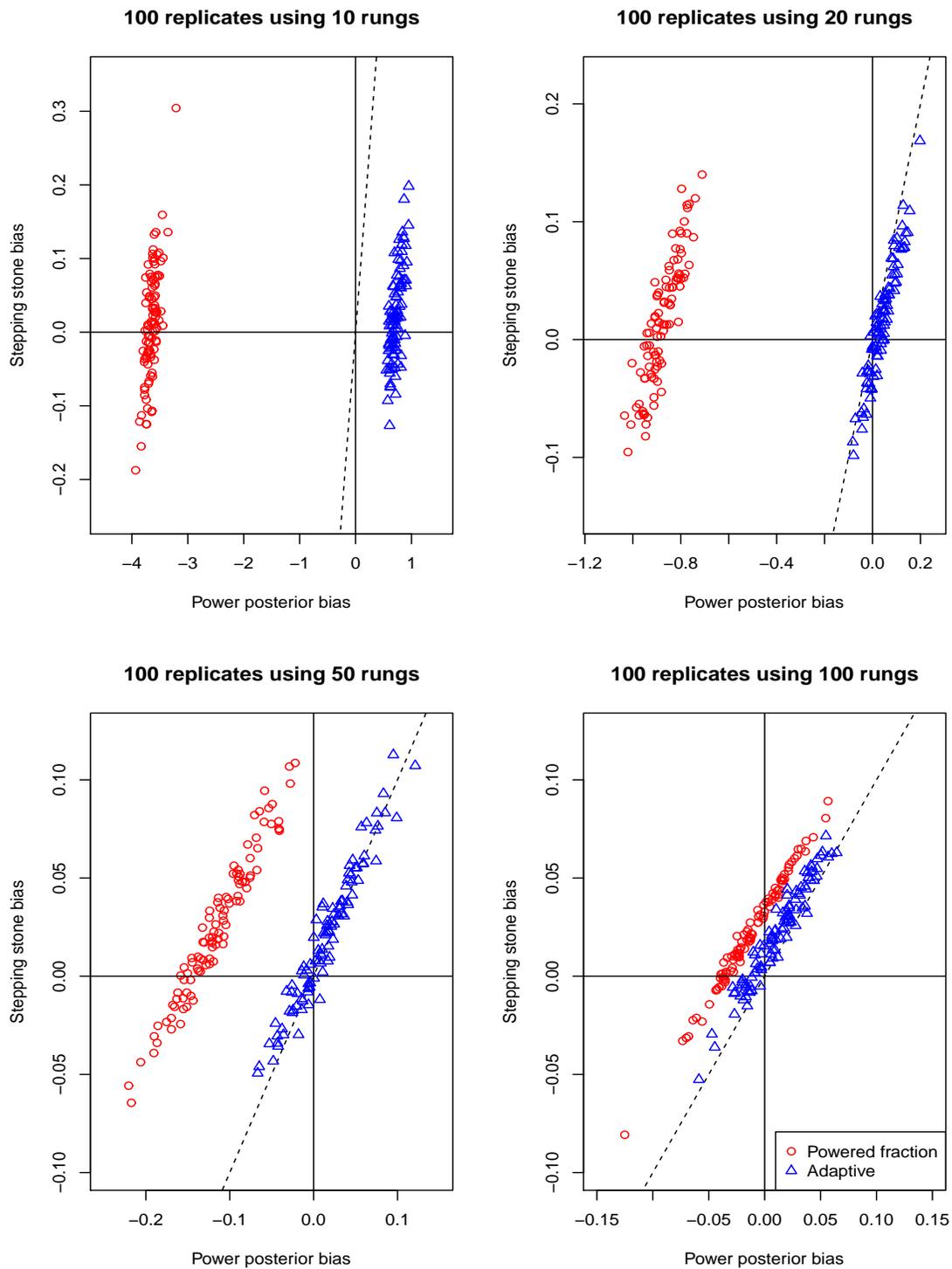}
\caption{Biases for Model 1 for the Pima data using the four methods and different numbers of rungs, matching up estimates calculated from the same long MCMC output (that is, power posteriors and Stepping Stones with PF spacing, and modified power posteriors and Stepping Stones with adaptive spacing). The solid black lines indicate zero bias, the dashed black lines represent the line $y=x$.}
\label{fig:pima2}
\end{center}
\end{figure}

Turning to the seemingly converged results in Table~\ref{tab:longer},
without any modification and with only PF spacing, Stepping Stones outperforms power posteriors, agreeing with the Xie {\it et al} findings.
In fact when $n=10$, even with modification, the bias associated with power posteriors dominates the RMSE (perhaps not surprisingly given the shape of the curve in the top panel of Figure~\ref{fig:pima1}).
For larger $n$, the modified power posterior with adaptive spacing generally does better in terms of RMSE than the PF Stepping Stones.
Introducing adaptive spacings to Stepping Stones does not greatly affect the bias, as expected, but it does reduce the standard error of the importance sampling based estimates.
Figure~\ref{fig:pima2} plots the corresponding Model 1 biases for the four methods and the different numbers of rungs, matching up estimates calculated from the same MCMC output (that is, power posteriors and Stepping Stones with PF spacing, modified power posteriors and Stepping Stones with adaptive spacing).
The most arresting feature here is perhaps the high degree of correlation between the estimates for all values of $n$.

\subsection{Example 3: Galaxy data}
\label{sec:galaxy}

To demonstrate a large application with a more challenging integral than the previous two, we use the much-studied Galaxy data set, see for 
example Richardson and Green (1997)\nocite{RichardsonBayesian}, which comprises
measurements on the velocities of 82 galaxies.
Denoting the 82 measurements by $y = \{y_1,\ldots,y_{82}\}$, 
we follow Richardson and Green (1997)\nocite{RichardsonBayesian}
in incorporating corresponding latent allocation variables 
$z = \{z_1,\ldots,z_{82}\}$.
Given $z_i=j$, $y_i$ follows the $j^{th}$ of the $k$ component Gaussian 
distributions of the mixture,
\begin{equation}
p( y_i | z_i=j , \mu_j,\sigma^2_j ) = \frac{1}
{\sqrt{2 \pi \sigma^2_j}} \exp
\left ( \frac{-(y_i-\mu_j)^2}{2 \sigma^2_j} \right ) \ \ \ i=1,\ldots,82.
\end{equation}
Conditional independence is assumed for the $\{y_i\}$ and
we specify independent standard proper priors:
\begin{eqnarray}
\mathsf{P}( z_i = j ) & = & w_j,\ \ \ \mbox{where } \sum_{j=1}^k w_j = 1 \\
\{ w_1 , \ldots , w_k \} & \sim & Dirichlet(1,k) \\
\mu_j & \sim & N( 0 , 1000 ), \ \ \ j=1, \ldots, k \\
\sigma^2_j & \sim & InvGam(1,1), \ \ \ j=1, \ldots, k. 
\label{eqn:3prior}
\end{eqnarray}
The weights, means and variances are all updated using the Gibbs sampler but we use a Metropolis algorithm with a discrete uniform proposal for the allocation variables.

\begin{figure}[tbp] 
\begin{center}
\includegraphics[width=15cm, height=20cm]{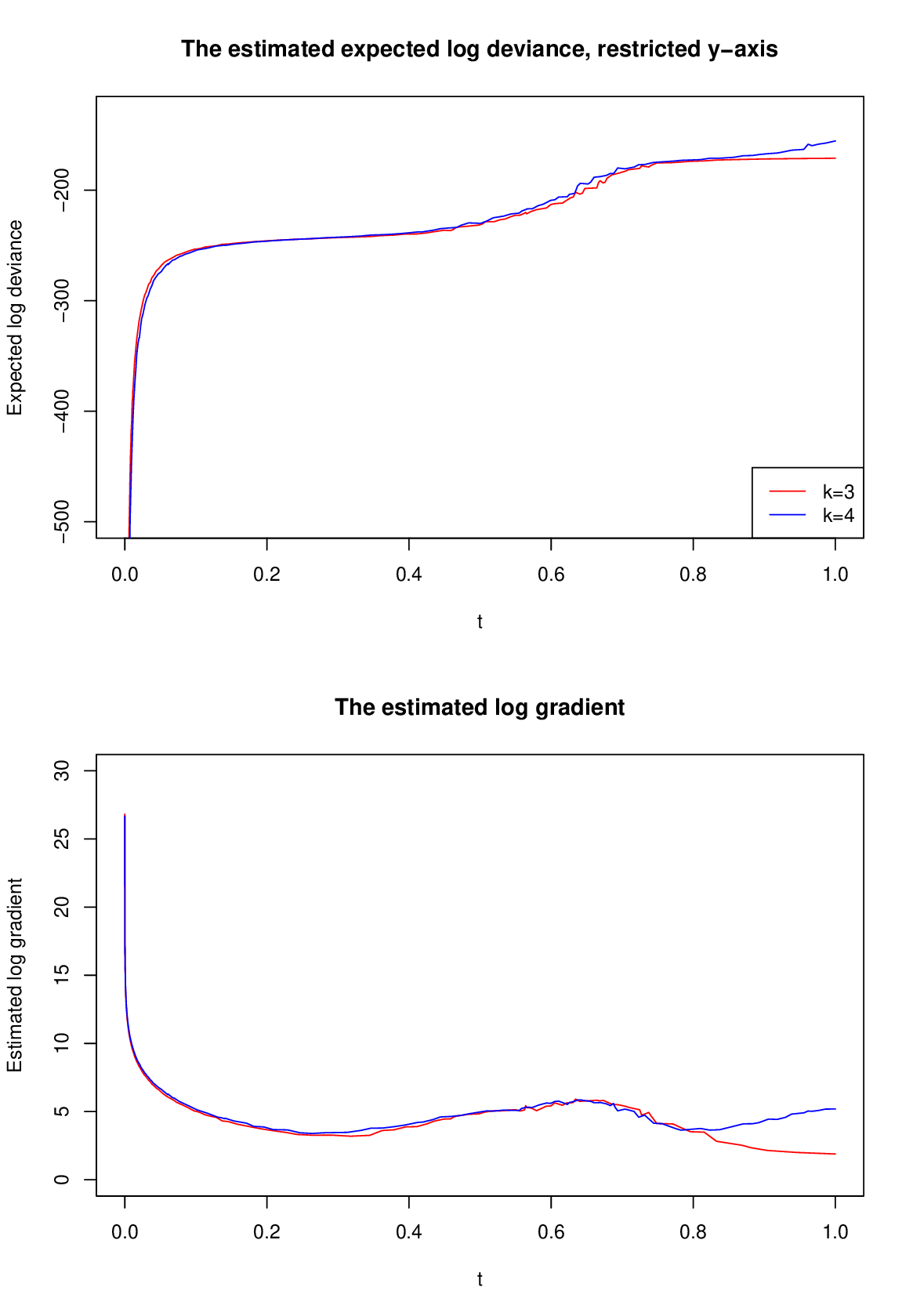}
\caption{Top: The expected log deviance curves for the Galaxy models using 200 rungs in the modified power posterior each with 50000 iterations, discarding the first tenth as burn in (plotted on a restricted vertical scale).
Bottom: The log of the corresponding estimated gradients.}
\label{fig:gal1}
\end{center}
\end{figure}

\begin{figure}[tbp] 
\begin{center}
\includegraphics[width=15cm, height=20cm]{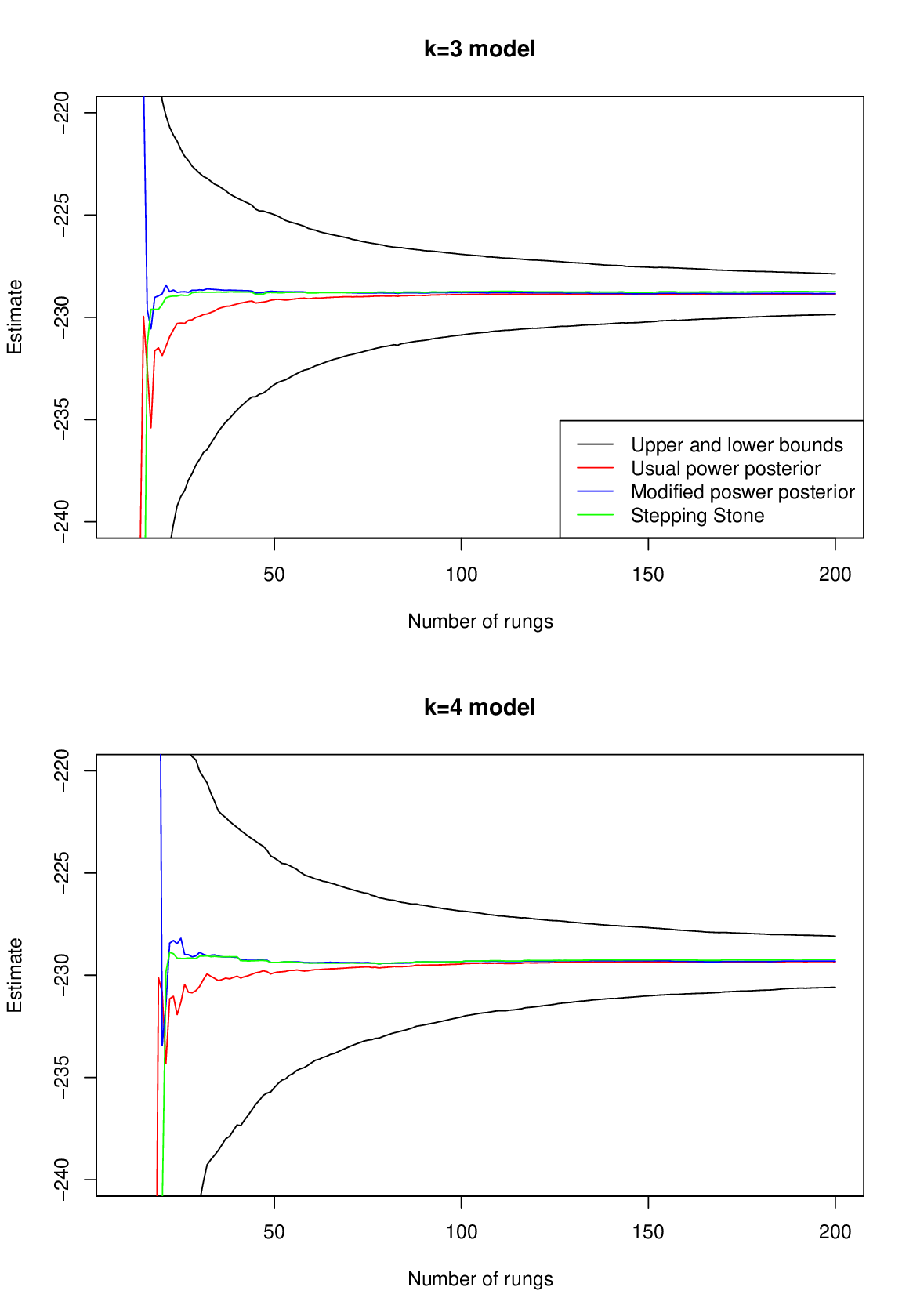}
\caption{Estimates of the log evidence for the Galaxy data based on the estimated curves and gradients in Figure~\ref{fig:gal1} for both $k=3$ and $k=4$ models, plotted for $n\ge10$.
Th e $y$-axes have been restricted to improve clarity.}
\label{fig:gal2}
\end{center}
\end{figure}

Behrens, Friel and Hurn (2012)\nocite{behrens2012tuning} considered the Galaxy data set and the above model when studying inverse temperature placement for the tempered transition algorithm, finding that its expected log deviance curve had some interesting features.
The top panel of Figure~\ref{fig:gal1} shows the estimated log deviance curves for $k=3$ and $k=4$.
These have been plotted on a restricted vertical scale so that interesting differences can be seen for larger inverse temperature values (the functions are in the region of -430000 when $t=0$).
These curves have been estimated using 200 adaptively placed inverse temperatures, with 50000 MCMC iterations at each rung, discarding the first tenth as burn in.
The sheer scale of the steepness towards zero is seen in the bottom panel of Figure~\ref{fig:gal1} which plots the corresponding estimated gradients on a log scale.
Based on these simulations, Figure~\ref{fig:gal2} gives the usual and modified power posterior estimates along with the Stepping Stone estimate.
It also plots an estimated upper and lower bound on the log evidence derived as the lower and upper step functions in the trapezium rule, Equation~(\ref{eqn:quadrature}).
The adaptive scheme has the benefit that if these estimated upper and lower discretisation bounds are still considered too wide after using the anticipated maximum number of inverse temperatures $n$, the process can simply be run on with additional inverse temperatures placed by the same algorithm (the same cannot be said of the PF scheme where it is not immediately clear how to add additional inverse temperatures).
In this case, for $k=3$, the modified power posterior estimate is -228.8428, the Stepping Stone estimate is -228.7486 with an estimated discretisation interval of $[-229.8626,-227.8737]$.
The corresponding figures for $k=4$ are -229.3113, -229.2291 and $[-230.5911,-228.0902]$.

\begin{figure}[tbp] 
\begin{center}
\includegraphics[width=15cm, height=20cm]{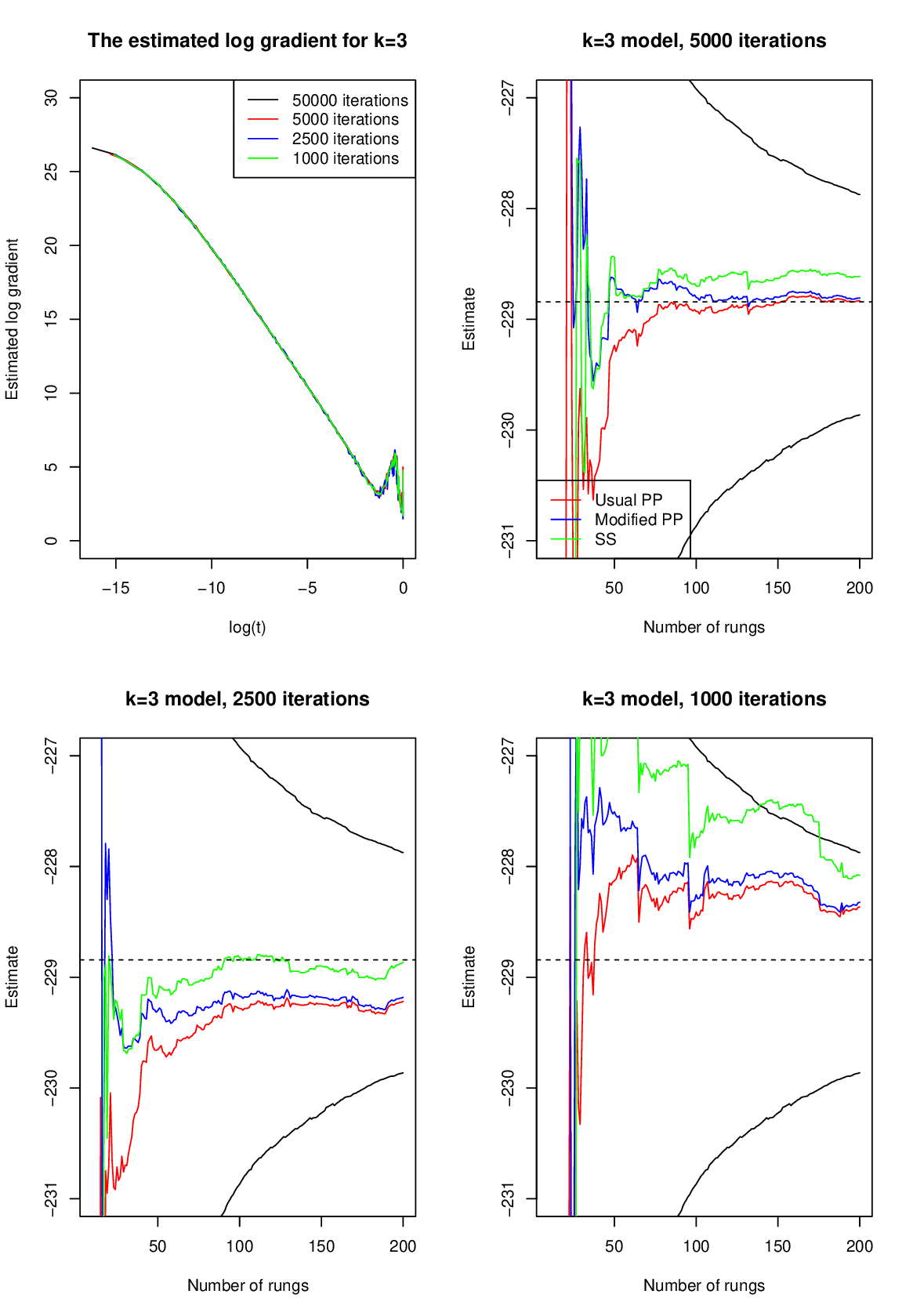}
\caption{Top left: The estimated log gradients against the log of the adaptive inverse temperatures for $k=3$.
Remaining three panels: the usual and modified power posterior estimates and Stepping Stone estimates plotted against the number of rungs. In all three cases, the black solid lines and black dashed line represent the estimated upper and lower bounds and final modified power posterior estimate from the $50000$ iteration run.}
\label{fig:gal3}
\end{center}
\end{figure}

\begin{figure}[tbp] 
\begin{center}
\includegraphics[width=15cm, height=20cm]{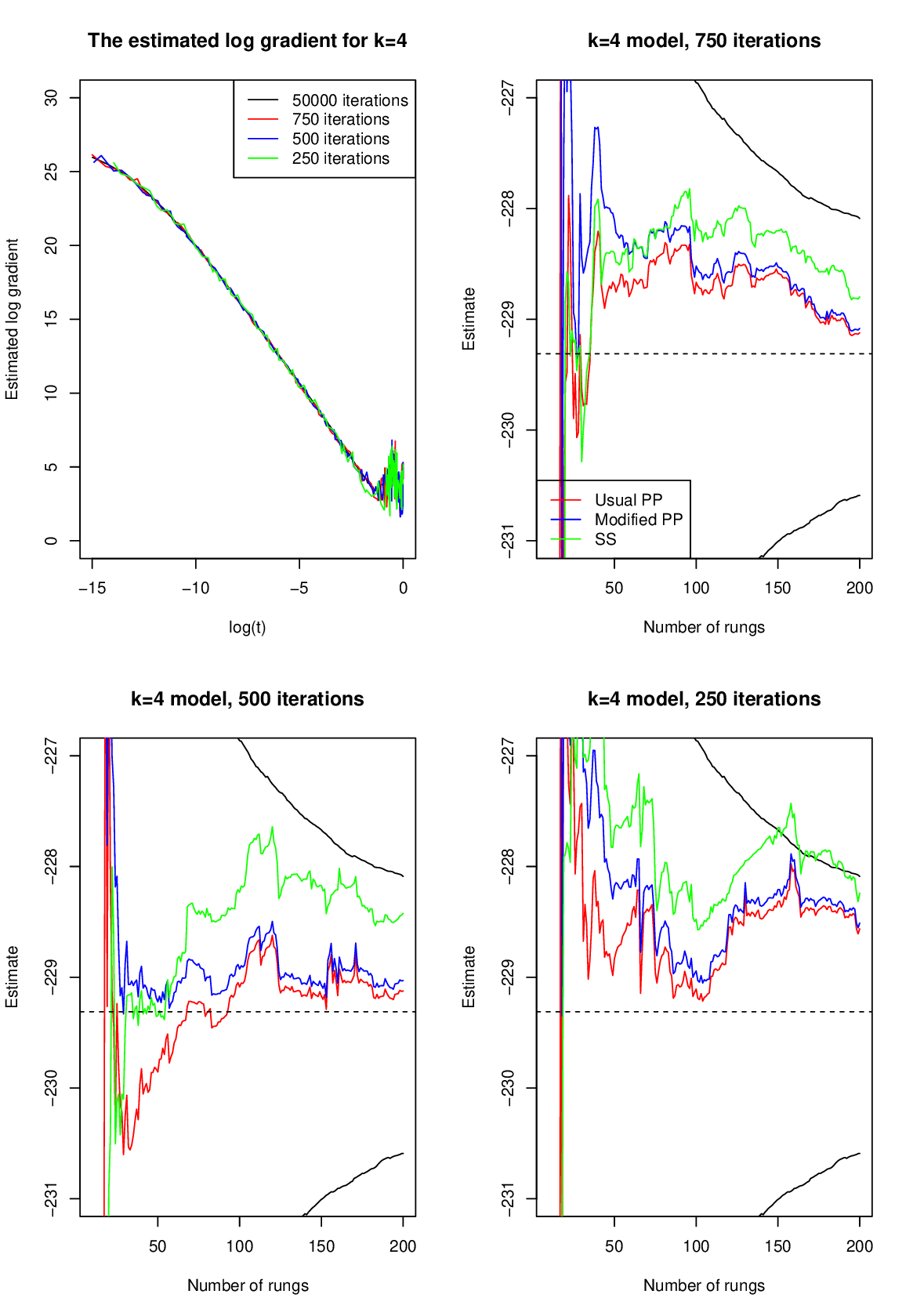}
\caption{Top left: The estimated log gradients against the log of the adaptive inverse temperatures for $k=3$.
Remaining three panels: the usual and modified power posterior estimates and Stepping Stone estimates plotted against the number of rungs. In all three cases, the black solid lines and black dashed line represent the estimated upper and lower bounds and final modified power posterior estimate from the $50000$ iteration run.}
\label{fig:gal4}
\end{center}
\end{figure}

The main aim of this section is to explore the effect uncertainty in estimating the gradients has on modified power posteriors using adaptive inverse temperature placements.
To do this, we consider progressively shorter runs of MCMC at each rung using the 50000 iteration run as a benchmark and retaining the burn in rule of the first tenth of the run.
Figure~\ref{fig:gal3} plots the log of the estimated gradient, Equation~(\ref{eqn:gradvar}), the usual and modified power posterior estimates and the Stepping Stone estimates for $5000$, $2500$ and $1000$ MCMC iterations for the $k=3$ model.
Figure~\ref{fig:gal4} plots the same quantities but now for $k=4$ and the rather short MCMC runs of $750$, $500$ and $250$ iterations.
Since most of the inverse temperatures are close to zero, we plot the log variance against the log of $t_i$ (which does mean we do not see the effect at $t=0$). 
Considering first Figure~\ref{fig:gal3}, reducing the number of MCMC iterations at each rung to 5000 or 2500 does not greatly affect the results with all three estimates lying within the upper and lower bounds, albeit not as tightly converging to the 50000 iteration final estimate.
As the run length decreases to 1000 or less (continuing on to Figure~\ref{fig:gal4}), we begin to see some discrepancies in the log variance curve and in the estimates themselves. 
Since we know that the effects of the adaptive placements and of the modification reduce as the number of rungs increases, we focus on $n \le 100$ (bearing in mind that reducing the run length also affects the estimated log deviance values themselves). 
The gaps between the three estimates grow (we are particularly interested in the gap between standard and modified power posteriors).
The Stepping Stone estimate does occasionally stray outside the lower and upper bounds, and there is again some evidence that it is veering towards higher values than the power posteriors.
The observation that when there are 50000 iterations the Stepping Stones and power posterior estimates are closer together than in these figures adds some more weight to our assertion that Stepping Stones is quite sensitive to poor convergence.  
However, although there is some indication that the estimates have not stabilised by the $n=200$ point, there are no catastrophic failures, just greater uncertainty.
Given that power posteriors rely on good estimates of of log deviance as well as its variance, it may be better to use a moderate size of $n$ with long MCMC runs at each $t_i$, rather than dividing up the same total number of iterations into short runs with a large $n$.

\section{Conclusions}
\label{sec:conclusions}

This article has, we hope, illustrated the potential gains that can be made when estimating the evidence using power posteriors by
correcting the numerical integration error and by adaptively choosing the inverse temperature ladder.
The methods that we have outlined
come at virtually no extra computational cost, and we would therefore recommend that these are routinely used when implementing
the power posterior approach. 

What this article does not do is to give guidance as to how to allocate computational resources between the different inverse temperatures.
We have seen in our examples that the gradient and thus also the variance of $\mathbf{E}_{\theta|y,t}\log(p(y|\theta))$ is largest as $t \rightarrow 0$, suggesting that we should allocate more MCMC iterations here rather than as $t \rightarrow 1$ to get good estimates.
On the other hand, when $t$ is small, the power posteriors $p_t(\theta|y)$ will probably be easy to sample compared to when $t=1$ and so we should also take into account the MCMC effective samples sizes.
Neither the gradients nor the effective sample sizes can be known before sampling is carried out! 
There probably is some scope for an adaptation scheme here too, perhaps allocating some fraction of the total number of MCMC iterations evenly over the inverse temperatures before allocating the remainder based on what we have learned in this initial phase.
This point also reinforces our caveat: what we have addressed here is discretisation error, the bounds we give are (noisy) bounds on this error and not credible intervals in the usual sense.

In our examples, applying the correction term has effectively smoothed over the benefits of the adaptive scheme over the PF one (just working a little harder in the latter case).
This numerical analysis trick is peculiar to this particular use of tempered distributions.
In general we suspect though that the adaptation ideas developed here might find wider use in other tempered schemes described in
the literature.

\paragraph*{Acknowledgements:} Nial Friel's research was supported by a Science Foundation Ireland Research Frontiers Program grant, 09/RFP/MTH2199. Jason Wyse's research was supported through the STATICA project, a Principal Investigator program of Science Foundation Ireland, 08/IN.1/I1879.
We are grateful to two anonymous reviewers whose comments on an earlier version have much improved this work.

\bibliographystyle{plain}
\bibliography{resultsBibliography}

\end{document}